\documentclass{elsart4-1}

\usepackage{graphicx}
\usepackage{subfigure}
\usepackage{color}
\usepackage{amsmath}
\usepackage{amssymb}
\usepackage[latin1]{inputenc}

\usepackage[english]{babel}

\newtheorem{e-proposition}[theorem]{Proposition}

\newtheorem{e-definition}[theorem]{Definition\rm}

\renewcommand{\theequation}{\arabic{equation}}
\newcommand{\be}{\begin{equation}}
\newcommand{\ee}{\end{equation}}
\begin{document}

\centerline{Astrophysics}
\begin{frontmatter}


\selectlanguage{english}
\title{On the cosmic ray spectrum from type II Supernovae expanding in their red giant presupernova wind}


\selectlanguage{english}
\author[blasi1]{Martina Cardillo},
\ead{martina@arcetri.astro.it}
\author[blasi1]{Elena Amato}
\ead{amato@arcetri.astro.it}
\author[blasi1,blasi2]{Pasquale Blasi},
\ead{blasi@arcetri.astro.it}

\address[blasi1]{INAF/Osservatorio Astrofisico di Arcetri, Largo E. Fermi, 5 - 50125 Firenze, Italy}
\address[blasi2]{Gran Sasso Science Institute (INFN), viale F. Crispi 7, 67100 L' Aquila, Italy}

\medskip
\begin{center}
{\small draft 1}
\end{center}

\begin{abstract}
While from the energetic point of view supernova remnants are viable sources of Galactic cosmic rays (CRs), the issue of whether they can accelerate protons up to a few PeV remains unsolved. Here we discuss particle acceleration at the forward shock of supernovae, and discuss the possibility that the current of escaping particles may excite a non-resonant instability that in turn leads to the formation of resonant modes that confine particles close to the shock, thereby increasing the maximum energy. This mechanism is at work throughout the expansion of the supernova explosion, from the ejecta dominated (ED) phase to the Sedov-Taylor (ST) phase. The transition from one stage to the other reflects in a break in the spectrum of injected particles. Because of their higher explosion rate, we focus our work on type II SNe expanding in the slow, dense wind, produced by the red super-giant progenitor stars. When the explosion occurs in such winds, the transition between the ED and the ST phase is likely to take place within a few tens of years. The highest energies are reached at even earlier times, when, however, a small fraction of the mass of ejecta has been processed. As a result, the spectrum of accelerated particles shows a break in the slope, at an energy that is the maximum energy ($E_{M}$) achieved at the beginning of the ST phase. Above this characteristic energy, the spectrum becomes steeper but remains a power law rather than developing an exponential cutoff. An exponential cut is eventually present at much higher energies but it does not have a phenomenological relevance. We show that for parameters typical of type II supernovae, $E_{M}$ for protons can easily reach values in the PeV range, confirming that type II SNRs are the best candidate sources for CRs at the \textit{knee}. 

From the point of view of implications of this scenario on the measured particle spectra, we have tried to fit KASCADE-Grande, ARGO -YBJ and YAC1-Tibet Array data with our model but we could not find any combination of the parameters that could explain all data sets. Indeed the recent measurement of the proton and helium spectra in the \textit{knee} region, with the ARGO-YBJ and YAC1-Tibet Array, has made the situation very confused. These measurements suggest that the \textit{knee} in the light component is at $\sim 650$ TeV, appreciably below the knee in the overall spectrum. On one hand this finding would resolve the problem of reaching very high energies in supernovae, but on the other it would open a critical issue in the transition region between Galactic and extragalactic CRs. 
\end{abstract}
\end{frontmatter}
\section{Introduction}
\label{sec:intro}
Supernova remnants (SNRs) can provide the correct order of magnitude of energy injection in the form of accelerated particles to explain the density of CRs observed in the Galaxy (for reviews see, e.g., \cite{blasi13rev,amato14rev}). However, the issue of whether particles can be accelerated diffusively at the SN shock to sufficiently high energies is still open for debate. One thing that is clear is that effective magnetic field amplification is necessary, since in the absence of this phenomenon the maximum energy is bound to be in the $\sim$ GeV region. This need was already recognized by \cite{lagage83a,lagage83b}, who found that the resonant instability induced by CR streaming could amplify the magnetic field to $\delta B/B\sim 1$ and allow the maximum energy to reach $\sim 10^{3}-10^{4}$ GeV. This latter upper bound on the maximum energy, still well below the \textit{knee}, derives from the limit $\delta B/B\lesssim 1$, which is in turn related to the resonant nature of the instability considered \cite{kulsrud69,wentzel74,skilling75a,skilling75b,skilling75c}. Non resonant instabilities may in fact potentially lead to larger amplification factors, but on scales either much larger or much smaller than the particle gyroradius, so that larger fields would not necessarily reflect into increased particle scattering, which is what is needed to reduce the acceleration time and increase the maximum achievable energy. 

The pioneering papers of \cite{lucek00} and \cite{bell04} showed how non-resonant modes can grow very fast ahead of a SNR shock. The fastest growing mode is characterized by wavenumber $k$ much larger than the Larmor radius of the particles contributing the CR current, but the non-linear evolution of these modes is shown to form fluctuations on larger scales, up to the Larmor radius of the particles constituting the CR current. At that point, resonant scattering may in principle facilitate the return of particles to the shock surface and possibly increase the maximum energy. It has been argued that this non-resonant branch of the streaming instability may account for very high energy CRs in the Galaxy \cite{bell13,schure13} and, in particular, may allow to reach the \textit{knee} if applied to the case of supernovae expanding in the wind of their red super-giant (RSG) progenitor star. The role of winds of evolved pre-supernova stars for CR acceleration has been discussed several times in the literature, with different spins. Since massive stars often go through a Wolf-Rayet (WR) evolutionary phase and they are usually clustered within OB associations (superbubbles), several authors (see \cite{binns08} for a recent review) have proposed that the collective effect of the WR fast rarefied winds may help accelerating CRs to very high energies and at the same time provide an explanation for some anomalies in the CR composition \cite{higdon98}. The most notable of such anomalies, and the one that is most solidly characterized from the observational point of view, is the ratio of abundances of two Ne isotopes, ${^{22}Ne}/{^{20}Ne}$ \cite{higdon03}, as observed in CRs, in comparison with its value in the solar environment. Additional constraints come from observations of $^{59}Ni$ in CRs, which impose limits on the rate of SN explosions in superbubbles, or, to state it more accurately, on the time when CR acceleration starts as compared with the time between two SN explosions in the same OB association. While this latter constraint is not a serious challenge for current models \cite{binns08}, it was recently pointed out \cite{prantzos12} that the average ${^{22}Ne}/{^{20}Ne}$ in superbubbles must approach the solar value, which would kill one of the main arguments in support of super bubbles as the main GCR sources.

In this paper we focus on the explosion of SNe in the wind of the RSG progenitor star and we argue that these explosions are the best sites to reach PeV energies. A number of SNe expanding in their RSG wind have been detected, among which the best studied case is that of SN 1993J \cite{ripero93}. This source is especially interesting for our purposes because radio measurements provide us with direct estimates of the magnetic field strength at different times, through the detection of the synchrotron self-absorption feature \cite{Fransson98}. This suggests a scenario in which not only the field is amplified at the SN shock, but the amplification turns out to be consistent with the expectations of the non-resonant instability \cite{Tatischeff09}. Because of the high density of such winds, the highest energies are reached within a few decades after the SN explosion, namely before the beginning of the ST phase of the explosion. In this scenario, particles are accelerated from a pool that is mainly made of wind particles, rather than ejecta of other SN explosions. Eventually one may expect that the shock reaches the edge of the bubble excavated by the stellar wind, and at that point particle acceleration proceeds in the ordinary way, although the details will depend on whether the star was originally in an OB association or in the ISM. At such late times, the particles escaping the SN are less energetic and we will not be concerned much with this stage of the acceleration process. On other hand, it is worth recalling that most constraints on composition and composition anomalies are limited to such lower energies.

The origin of the \textit{knee} in the all-particle spectrum of CRs is inextricably connected with the issue of the end of the Galactic CR spectrum and the transition to extragalactic CRs: in Ref.~\cite{aloisio14}, the authors try to model the transition region by summing the Galactic SNR contribution and the flux of nuclei of extragalactic origin, required to fit the Auger data \cite{auger10}. The authors reach the conclusion that in order to satisfy the observational requirement of a predominantly light composition in the EeV region \cite{auger10,HiREs04,HiREs08,TA11}, an extra component of extragalactic protons is required. Such an extra component appears to be in good agreement with the proton spectrum as measured by KASCADE-Grande \cite{apel13}. However, both the proton and iron spectra measured by KASCADE-Grande in the energy region $10^{16}-10^{18}$ eV suggest that either the injection spectra are not cut off exponentially at the maximum energy (namely the number of particles decreases more slowly than an exponential at energies $E\gg E_{M}$), or there is some, as yet unknown, class of sources with maximum energy much in excess of the \textit{knee} energy. 

The observational situation has recently become even more confusing after the publication of the spectra of light CRs (protons and helium) by the ARGO-YBJ \cite{argo} and YAC1-Tibet Array \cite{yac} experiments. Both measurements suggest a \textit{knee} in the light component at $\sim 650$ TeV, appreciably below the \textit{knee}. On one hand, it is clear that this finding would make the requirements in terms of particle acceleration less severe. On the other, however, it makes the requirements on the transition region much harder to accommodate. In other words, even in the scenario arising from ARGO data a population of sources able to reach maximum energy much above the \textit{knee} is required to exist, thereby leaving the problem of the maximum energy little affected.  

In this paper we consider the viability of the two hypotheses above: we investigate whether non-exponential tails at $E>E_{M}$ are to be expected in the spectra of particles escaping supernova shocks, and we study the possibility of having supernovae that contribute maximum (proton) energies much higher than the \textit{knee}. The paper is organised as follows: in \S~\ref{sec:bell} we discuss the role of escaping particles for the estimate of the maximum energy achievable at a supernova shock. In \S~\ref{sec:NHR} we illustrate our calculations of the spectra of nuclei and in \S~\ref{sec:results} we compare our predictions with the observed spectra, with particular emphasis on the \textit{knee} and the transition region. In \S~\ref{sec:data} we discuss the implications of the recent measurements of the ARGO-YBJ and YAC1-Tibet Array, comparing them with KASCADE-Grande data. We conclude in \S~\ref{sec:summary}. 

\section{Non-resonant instability excited by escaping CRs}
\label{sec:bell}

CR streaming leads to the excitation of both resonant Alfv\'en waves and non-resonant, almost-purely-growing, modes \cite{bell04}. The former have a growth rate that, for typical parameters of a supernova remnant, leads to maximum energy $\ll PeV$, mainly as a consequence of the saturation at $\delta B/B\sim 1$. On the other hand, the non-resonant instability is very fast growing, but the fastest growing mode has a wavelength much shorter than the Larmor radius of the particles generating the current, and also the wrong polarisation for resonant interaction with positively charged particles moving away from the shock surface \cite{amatoblasi09}. As a result, to first order, the non-resonant waves do not lead to efficient particle scattering, and would therefore be useless to the goal of increasing the maximum achievable energy.

However, it was already suggested by \cite{bell04}, and then confirmed by the numerical work of \cite{riquelme09} and more recently \cite{caprioli14}, that the non-linear evolution of these modes leads to the generation of power on larger spatial scales. Here it is useful to think in physical terms about the process that may be responsible for particle return to the shock from the upstream region. Let us consider a first generation of particles of energy equal to the current achievable maximum, $E$, moving from downstream to upstream. By definition, upstream there are no waves able to scatter resonantly such particles, hence they are going to move quasi-ballistically and escape the system at the speed of light (or fractions of it, depending on the level of anisotropy of the distribution of escaping particles).
The current of particles at distance $R$ from the explosion center as due to particles escaped when the shock location was at $r$ can be written as:
\begin{equation}
j_{CR}(R,r)=n_{\rm CR,r}\left(R,E_M(r)\right) e v_{sh}(r) =e \frac{\xi_{CR} \rho(r) v_s^3}{E_0\Psi(E_M)}\left(\frac{r}{R}\right)^2
\label{eq:jCR}
\end{equation}
with
\be
\Psi=\left\{
\begin{array}{ll}
\left(E_M/E_0\right)\ln \left(E_M/E_0\right) & \beta=0\\
 & \\
\frac{1+\beta}{\beta}\left(\frac{E_M}{E_0}\right)^{1+\beta}\left[1-\left(\frac{E_0}{E_M}\right)^\beta\right] & \beta\ne0\ .
\end{array}
\right.
\label{eq:psi}
\ee
In the above expressions we have used the relation between the number density of accelerated particles with energy $E$, $n_{CR}(E)$, at some location $R$ upstream of the shock, and the energy density in accelerated particles at the shock $\xi_{CR} \rho(r)v_s^2(r)$, where $\xi_{CR}$ is the CR acceleration efficiency. This relation is easily found through the transport equation for accelerated particles at a plane shock. The function $\Psi(E_M)$ accounts for normalisation of the particle distribution function which is taken to be $f_s(E)\propto E^{-(2+\beta)}$ and extending between a minimum energy $E_0$, that does not depend on time and a maximum energy $E_M$ which does depend on time, as we will see.
In Eq. \ref{eq:jCR}, the factor $(r/R)^{2}$ should account for the dilution due to spherical propagation of particles in ballistic motion. 
The underlying assumption is that the large scale magnetic field is radial or absent. Whether this assumption is a realistic one in the two environments we consider in the following, namely the ISM or the wind of a massive progenitor star, may be questionable. In the latter case the zeroth order expectation is that the field should be toroidal, although one can envision that instabilities induced by the interaction between the slow and dense wind produced during the red supergiant phase of the progenitor and the faster and more dilute wind later blown as a Wolf Rayet star, could stretch the field lines and make them radial in at least a portion of the shock surface. In the case of a SNR expanding in the ISM, more caution should be taken, in that there is no reason why the upstream magnetic field should display a radial geometry.

The fastest growing mode has a growth rate
\begin{equation}
\gamma_{M} = k_{M} v_{A},
\label{eq:gmax}
\end{equation}
where $v_{A}$ is the Alfv\'en speed in the unperturbed magnetic field $B_{0}$. The wavenumber where the growth is the fastest can also be easily estimated using the condition 
\begin{equation}
k_{M} B_{0}\cong\frac{4\pi}{c}j_{CR},
\label{eq:Btension}
\end{equation}
which corresponds to balance between current and magnetic tension. The above expression implies that:
\begin{equation}
k_{M}r_{L}=\xi_{CR}\frac{1}{\Psi}\left(\frac{v_{sh}}{V_{A}}\right)^{2}\left(\frac{v_{sh}}{c}\right)\gg 1,
\label{eq:non_resonance}
\end{equation}
where $r_{L}$ is the particle gyroradius, $\xi_{CR}$ is the CR efficiency and $V_{A}=\frac{B_{0}}{\sqrt{4\pi \rho}}$ the Alfv\'en velocity. 
Upon saturation of the instability, the energy density in the form of amplified magnetic field in units of the ram pressure of the plasma can be estimated as $(v_{sh}/c)(\xi_{CR}/\ln(E_{M}/E_{0}))$, that for typical values of the parameters is $\sim 10^{-4}$ (for simplicity here we assumed that the spectrum of accelerated particles is $\sim E^{-2}$). This reflects in about one order of magnitude larger pressure downstream of the shock (as due to compression), an estimate which is still a factor of 10 below that of \cite{ptuskin10}, where the magnetic pressure downstream was assumed to be a fixed fraction ($\sim3.5\%$) of the ram pressure, independent of the shock speed.

Numerical simulations show that the saturation of the instability occurs after $N_t\sim 5$ e-folds \cite{bell04}, namely 
\begin{equation}
\int \gamma_{M}dt=\int (k_{M}v_{A}) dt \approx 5\ .
\label{eq:growthrate}
\end{equation}
Using the expression for $k_{M}$ derived in Eq.~\ref{eq:Btension}, we can rewrite this condition, at a point $R$ upstream of the shock, as \cite{bell13,schure13,schure14,blasi14}:
\begin{equation}
\int_{0}^{R} \frac{4 \pi e}{c E_0} \frac{\xi_{CR} \rho(r)v_{sh}(r)^{2}}{\sqrt{4\pi\rho(R)}\Psi(E_M(R))}\left(\frac{r}{R}\right)^{2}dr\sim 5\ ,
\label{eq:Emax_int}
\end{equation} 
where we have used the shock velocity to change the integration variable from time to shock coordinate.
In Eq. \ref{eq:Emax_int}, the integral in $r$ is assumed to extend over all previous radii of the shock. In fact, since the values of the growth rate that are found for typical values of SN parameters are rather large ($\gamma_{M}\gg v_{sh}/R$), in case of ballistic motion of the escaping particles, one would expect that only a narrow range of radii close to $R$ would give contribution to the particle current. Here we keep the form of Eq. \ref{eq:Emax_int} as given in Refs. \cite{bell13,schure13,schure14}, since the difference that this correction would introduce is only a factor of $\sim 2$.

In order to find an expression for the evolution of the maximum energy with time, we need to assume a density profile of the medium in which the supernova remnant is expanding. For a supernova expanding in the uniform ISM, one can assume that the ISM density is constant $\rho(R)=\rho_{ISM}$. For a type II supernova, instead, it is often the case that the explosion takes place in the wind produced by a red giant pre-supernova progenitor star. The density of the wind can be written as 
\begin{equation}
\rho(R)\cong\frac{\dot{M}}{4\pi R_{0}^{2}V_{w}}\left(\frac{R_0}{R}\right)^{2}=\rho(R_{0})\left(\frac{R_{0}}{R}\right)^{2},
\label{eq:density}
\end{equation}
where $\dot{M}$ is the rate of mass loss of the red giant and $V_{w}$ is the wind velocity. In the following we will simply write $\rho(R)\propto R^{-m}$, having in mind the two cases of $m=0$, corresponding to a uniform medium, and $m=2$, appropriate for expansion in the progenitor wind.

Differentiating Eq.~\ref{eq:Emax_int} with respect to $R$, we obtain an implicit expression for the maximum energy (see Fig.~\ref{fig:times1}):
\begin{equation}
\Psi(E_{M}(R))\cong\frac{2 e}{(4-m)5c E_0}\xi_{CR} v_s(R)^2 \sqrt{4 \pi \rho(R) R^2}\ .
\label{eq:Em_general}
\end{equation}

\section{From escape of accelerated particles to CRs}
\label{sec:NHR}
\subsection{Spectrum of the escaping particles}
\label{sec:escaping}

For a supernova shock moving with velocity $v_{sh}(t)$ in the surrounding medium (ISM or wind of the progenitor star) the shock radius is $R_{sh}(t)=\int_{0}^{t} dt' v_{sh}(t')$. At each time $t$ we assume that particles with given energy $E_M(t)\equiv E$ can escape the accelerator. 
The number of particles that escape the shock at each give time is related to the current at the upstream radius $R$ through:
\be
N_{esc}\left(E\right)dE=\frac{J_{CR}}{e} 4 \pi R^2 dt\ ,
\label{eq:nesc}
\ee
where $N_{esc}\left(E\right)$ is the number of particles per unit energy, with energy $E=E_M(t)$, that are released in the ISM. 
Using the expression in Eq.~\ref{eq:jCR} for the current one can rewrite:
\be
N_{esc}\left(E\right)=\frac{\xi_{CR} \rho v_s^2}{E_0 \Psi} 4 \pi R^2 \frac{dR}{dE}
\label{eq:nesc2}
\ee
with $dR/dE=(dR/d\Psi) (d\Psi/dE)$ to be computed using Eq.~\ref{eq:psi} and \ref{eq:Em_general}.
In order to derive the first term on the right in the latter equation, we need to make explicit assumptions on the shock dynamics. For purposes of illustration we assume in the following a general power-law dependence of the shock radius on time: $R\propto t^\lambda$, which also implies $v_s\propto R^{(\lambda-1)/\lambda}$.
Using the latter dependencies in Eq.~\ref{eq:Em_general} it is straightforward to find for $dR/d\Psi$:
\be
\frac{dR}{d\Psi}=\frac{R}{\Psi}\left(2 \frac{\lambda-1}{\lambda}-\frac{m}{2}+1\right)^{-1}=\frac{R}{\Psi}\frac{1}{S(\lambda,m)}
\label{eq:dpsidr}
\ee
and rewrite Eq.~\ref{eq:nesc2} in a particularly useful way:
\be
N_{esc}(E)=\frac{4 \pi \xi_{CR}}{E_0 S(\lambda,m)}\rho v_s^2 R^3\chi(E)\ ,
\label{eq:nescfin}
\ee
where $\chi(E)=(d/dE)(1/\Psi(E))$ is a function of $E$ alone, to be computed from Eq.~\ref{eq:psi}. 

The advantage of Eq.~\ref{eq:nescfin} is that it makes it immediately clear that in the Sedov-Taylor phase, during which $\rho v_s^2 R^3$ is constant with time, the CR spectrum released in the ISM is bound to be just given by $\chi(E)$, and hence only depend on the spectrum in the remnant, unless $\xi_{CR}$ or $E_0$ depend on time. It should also be noticed, however, that this dependence is not just a one to one identity and the spectrum of escaped CRs will not be the same as the spectrum of accelerated particles in the remnant, as already noted by \cite{bell13},\cite{caprioli14}. In fact, when computing $\chi(E)$ explicitly from Eq.~\ref{eq:psi} we find:
\be
\chi(E)\approx\frac{d}{dE}\left(\frac{1}{\Psi}\right)=\frac{1}{E_0}
\left\{
\begin{array}{ll}
\frac{\beta}{1+\beta} \left(\frac{E}{E_0}\right)^{-2} & \beta<0\\
&   \\
\left(\frac{E}{E_0}\right)^{-2}\left[\frac{\left(1+\ln(E/E_0)\right)}{(\ln(E/E_0))^{2}}\right] & \beta=0 \\
 & \\
\beta \left(\frac{E}{E_0}\right)^{-(2+\beta)} & \beta>0
\end{array}
\right.
\label{eq:dpside}
\ee
where the sign $\approx$ refers to the fact that we have used the assumption $E>>E_0$. It is apparent that the power-law dependence of $\chi(E)$ can only be $-2$ or steeper. Therefore the spectrum of CRs released during the Sedov-Taylor phase is the same as the spectrum of accelerated particles in the source only if the latter is $E^{-2}$ or steeper, while for flatter source spectra $N_{esc}(E)\propto E^{-2}$. 

While in the Sedov-Taylor phase, only the environmental details are important while the ejecta density profile is irrelevant to the spectral slope, during the free expansion phase both density profiles are important
in order to determine the position of the breaks and the maximum energy to which the released spectrum extends. This kind of information enters the calculation of the dependence of the shock radius on time, which we expressed as $R\propto t^\lambda$.

In the following, we take the value of $\lambda$ appropriate to describe the different evolutionary stages of the SNR from Ref.~\cite{chevalier89,ptuskin05,truelove99}. This work gives the remnant expansion law during the different stages for generic density profiles of both the SN ejecta $\rho_{ej}\propto R^{-k}$ and the ambient medium $\rho\propto R^{-m}$. The general expression for $\lambda$ is:
\be
\lambda=\left\{ \begin{array}{ll}
 \frac{k-3}{k-m}& \,\,\,\,\,\textrm{in the ED phase}\\
 \\
\frac{2}{5-m} & \,\,\,\,\,\textrm{in the ST phase}\ ,
\end{array} \right.
\ee
where the ejecta profile appropriate to describe the two types of SN has $k=7$ and $k=9$ for type I and II explosions respectively.

Once $\lambda$ is given, from Eq.~\ref{eq:nescfin} and \ref{eq:dpside} we can derive the spectrum released by the SN during the ED and ST phases as:
\be
N_{esc}(E)\propto \rho v_s^2 R^3 E^{-(2+\epsilon)}\propto E^{\frac{(1+\epsilon)(4\lambda-m\lambda)-(6\lambda-4-m\lambda)}{6\lambda-4-m\lambda}} 
\label{eq:nescprop}
\ee
with $\epsilon=0$ if $f_s(E)\propto E^{-(2+\beta)}$ with $\beta \leq0$ and $\epsilon=\beta$ if $\beta>0$.
Inverting the time dependence of $E$ as can be found from Eq.~\ref{eq:Em_general}, and expressing the dynamical quantities as functions of $E$, one finally finds: 
\be
N_{esc}(E)\propto 
\left\{ 
\begin{array}{ll}
E^{-(5+4\epsilon)}& \,\,\,\,\,\textrm{ED phase;}\\
 & \, \, \, \, \, \, \, \,\, \, \, \,\, \, \, \,\, \, \, \,\, \, \, \,m=0, k=7\\
E^{-(2+\epsilon)}& \,\,\,\,\,\textrm{ST phase;}
\end{array} 
\right.
\ee
\be
N_{esc}(E)\propto\left\{ 
\begin{array}{ll}
E^{-(4+3\epsilon)}& \,\,\,\,\,\textrm{ED phase;}\\
 & \, \, \, \,\, \, \, \,\, \, \, \,\, \, \, \,\, \, \, \,\, \, \, \,m=2, k=9\\
E^{-(2+\epsilon)}& \,\,\,\,\,\textrm{ST phase;}
\end{array} \right.
\label{eq:spec}
\ee
This result is very interesting because it shows that the spectrum of escaping CRs integrated over time during the SN shock expansion is a broken power law, with an index close to 2 at low energies (below$E_{M}$) and steeper at higher energies. In more general terms, the spectrum of CRs released into the ISM is different from the postshock spectrum of accelerated particles, and is rather sensitive to how magnetic field amplification evolves in time \cite{caprioli09,caprioli10}. A similar conclusion was previously obtained in the context of simple box models of CR acceleration at shocks \cite{Drury03,Drury11}. At some energy $\gg E_{M}$, the spectrum will eventually suffer an exponential cutoff. However such cutoff does not have an immediate phenomenological impact, because it would appear at energies at which the spectrum has already dropped appreciably due to the steeper power law. The transition between the two power laws at the maximum energy $E_{M}$ corresponds to the transition between the end of the ED phase, where most mass has already been processed, and the beginning  of the ST (adiabatic) phase. Notice that in principle the highest value reached by particles at the shock is larger than $E_{M}$ at earlier times, but the flux of CRs with such energies is suppressed (though not exponentially but as a power law) because of the small amount of mass that is processed during the ED phase.

In order to provide an estimate of the maximum achievable energy as a function of time, we proceed with modeling the transition from the ED to the ST phase of the SN expansion by using a parametric form for the time dependence of the shock radius:
\begin{equation}
R_{sh}(t)=R_{0}\left[\left(\frac{t}{t_{0}}\right)^{a\lambda_{ED}}+\left(\frac{t}{t_{0}}\right)^{a\lambda_{ST}}\right]^{1/a}
\label{eq:Rsh}
\end{equation}
where $a$ is a smoothing parameter (in the following we use $a=-5$ unless indicated otherwise), $\lambda_{ED}$ and $\lambda_{ST}$ are the indices describing the asymptotic time dependencies during the ED and ST phases, respectively, and $R_{0}$ and $t_{0}$ are the values of radius and time at the transition between the two phases. The value of $R_{0}$ can be estimated by using the condition that at such distance the swept up mass equals that of the ejected material $M_{ej}$:
\begin{equation}
M_{ej}=4\pi\int^{R_0}_{0}\rho(R) R^2 dR,
\label{eq:Mej}
\end{equation}   
where $\rho(R)$ is the ISM density in the case of type I SNe and the density in the wind in the case of type II explosions. For the former type we then find, for the transition radius: 
\be
R_0=\left(\frac{3 M_{ej}}{4 \pi \rho} \right)^{1/3}\approx 2\ \left(\frac{M_{ej}}{M_\odot}\right)^{1/3} \left(\frac{n_{ISM}}{\rm{cm^3}}\right)^{-1/3}\  pc\ 
\label{eq:R0I}
\ee
where $\rho$ is the constant ISM density, which we take to correspond to 1 particle ${\rm cm}^{-3}$.
For type II events, instead:
\begin{equation}
R_{0}=\frac{M_{ej}V_{w}}{\dot{M}}\approx 1\ \left(\frac{M_{ej}}{M_\odot}\right) \left(\frac{V_w}{10\,{\rm km\,s^{-1}}}\right) \left(\frac{\dot M}{10^{-5} M_\odot \rm{yr^{-1}}}\right)^{-1}\,pc\ ,
\label{eq:R0II}
\end{equation}

The assumption of self-similarity for the time evolution of the SN shock leads to a generic form for the density evolution of the ejecta \cite{truelove99}
\begin{equation}
\rho_{ej}(R,t)=A\times R^{-k}t^{k-3}
\label{eq:ejecta_density}
\end{equation}
where 
\begin{equation}
A=\frac{1}{4\pi k}\frac{\left[3(k-3)M_{ej}\right]^{5/2}}{\left[10(k-5)E_{SN}\right]^{3/2}}\left[\frac{10(k-5)E_{SN}}{3(k-3)M_{ej}}\right]^{k/2}\ ,
\end{equation}
and $E_{SN}$ is the SN explosion energy. 
The condition that the density of the ejecta and the density of the wind be equal at the forward shock leads to the following expression for the time $t_{0}$:
\begin{equation}
t_{0}=\left[R_{0}\left(\frac{B}{A}\right)^{1/(k-m)}\right]^{\frac{k-m}{k-3}}
\label{eq:t0}
\end{equation}
where $B=\rho_{ISM}$ for type I and $B=\dot{M}/(4\pi V_{w})$ for type II respectively.

The shock velocity can be easily written as
\begin{equation}
v_{sh}(t)=\frac{R_{0}}{t_{0}}\left(\frac{R}{R_{0}}\right)^{1-a}\left[\lambda_{ED}\left(\frac{t}{t_{0}}\right)^{a\lambda_{ED}-1}+\lambda_{ST}\left(\frac{t}{t_{0}}\right)^{a\lambda_{ST}-1}\right],
\label{eq:vsh}
\end{equation}
which can be used to calculate the maximum energy as a function of time, and hence the spectrum of escaping particles from an individual supernova.

Before proceeding with detailed calculations, let us briefly discuss our choice of focusing on type II SNe as the best candidate producers of CRs at the \textit{knee}. Let us consider the maximum energy that can be reached by each type of SN events, as given by Eq.~\ref{eq:Em_general} evaluated at the transition between the ED and the ST phase. To the goal of a simple estimate, let us specialise this equation to the case of a $E^{-2}$ source spectrum. 

For type I events we can estimate:
\be
E_{M}\cong\frac{2 e}{10c}\xi_{CR} v_{0}^2 \sqrt{4 \pi \rho R_{0}^2}=130\ \left(\frac{\xi_{CR}}{0.1}\right) \left(\frac{M_{ej}}{M_\odot}\right)^{-\frac{2}{3}} \left(\frac{E_{SN}}{10^{51}\rm{erg}}\right) \left(\frac{n_{ISM}}{\rm{cm^{-3}}}\right)^{\frac{1}{6}}\ TeV
\label{eq:emaxI}
\ee
while for type II:
\be
E_{M}\cong\frac{2 e}{5c}\xi_{CR} v_{0}^2 \sqrt{4 \pi \rho R_{0}^2}\approx1\ \left(\frac{\xi_{CR}}{0.1}\right)\left(\frac{M_{ej}}{M_\odot}\right)^{-1} \left(\frac{E_{SN}}{10^{51}\rm{erg}}\right) \left(\frac{\dot M}{10^{-5} M_\odot \rm{yr^{-1}}}\right)^{\frac{1}{2}} \left(\frac{V_w}{10\,\rm{km\, s^{-1}}}\right)^{-\frac{1}{2}}\ PeV\ .
\label{eq:emaxII}
\ee
It is apparent that for standard parameters pertaining the two types of explosions, type II SNe can reach about one order of magnitude larger maximum energies than type Ia, and in particular that if particles are accelerated and escape as described here, they easily seem to be able to provide proton acceleration up to the \textit{knee}. 

\begin{figure}[!h]
\centering
\includegraphics[scale=1]{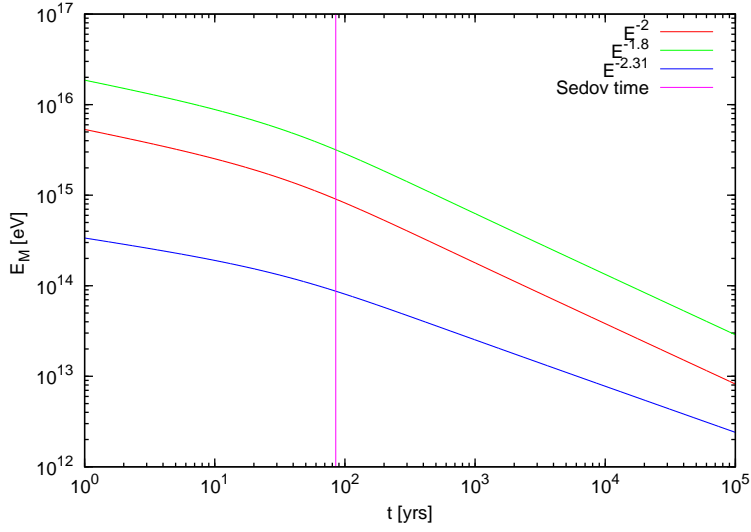}  
\caption{The maximum energy as a function of time after a type II SN explosion with $E_{SN}=10^{51}$ erg, $\xi_{CR}=0.1$, $M_{ej}=M_\odot$, $\dot M=10^{-5} M_\odot yr^{-1}$ and $V_w=10 km\ s^{-1}$. The different curves refer to different injection indices: $p=-2$ in red, $p=-1.8$ in green, $p=-2.31$. The vertical line identifies the time of beginning of the ST phase}.
\label{fig:times1}
\end{figure}

In order to illustrate how the maximum energy depends on the assumed spectral index within this framework, in Fig.~\ref{fig:times1} we plot the evolution with time of the instantaneous maximum energy a type II SNR can achieve for three different values of the source spectral index (all the other parameters have standard values, as specified in the caption). The figure shows that $E_M$ as derived from Eq.~\ref{eq:Em_general} and \ref{eq:psi} changes by about 1 order of magnitude at the transition between ejecta dominated and Sedov phase (indicated by the vertical line in the plot) depending on the source spectral index: spectral indices steeper than 2 lead to lower values of $E_M$ compared to the estimate in Eq.~\ref{eq:emaxII}, while the opposite is true for flatter spectra.

\subsection{CR Spectra observed at the Earth}
\label{sec:observed}

Here we adopt a simple model for the description of the transport of nuclei in the Galaxy. Assuming that the transport has reached a stationary regime and that it is dominated by diffusion, the equation describing the transport of a nuclear specie $\alpha$ is:

\begin{equation}
\frac{\partial}{\partial z}\left(D_{\alpha}\frac{\partial f_{\alpha}}{\partial z}\right)+Q_{\alpha}- f_{\alpha} n \sigma_{sp,\alpha} v = 0,
\label{eq:transport}
\end{equation} 
where $Q_{\alpha}$ is the rate of injection per unit volume of CR of type $\alpha$ and $\sigma_{sp,\alpha}$ is the cross section for spallation. In principle one should include an injection term of particles of type $\alpha$ due to the spallation of heavier nuclei, but as long as we focus on primary species and we are at sufficiently high energies, this contribution can be neglected \cite{berezinskii90}. Eq.~\ref{eq:transport} can be easily solved if we assume that both injection and spallation occur in a thin disc of size $2h$:
\begin{equation}
n(z) = n_{d} 2 h \delta (z) ~~~~~ Q_{\alpha}(p) = \frac{N_{\alpha}(p) \Re}{\pi R_{D}^{2}} \delta(z).
\end{equation}
Here the injection spectra $N_{\alpha}(p)$ are calculated as discussed in Sec.~\ref{sec:escaping}, $n_{d}$ is the gas density of the Galactic disc, assumed of thickness $2h$ and radius $R_{D}$. If we introduce the grammage
\begin{equation}
X(p) = n_{d} \frac{h}{H} v m_{p} \frac{H^{2}}{D(p)},
\end{equation}
and we impose a free escape boundary condition at $|z|=H$, the solution of Eq. \ref{eq:transport} in the Galactic disc can be easily written as follows:
\begin{equation}
f_{0,\alpha}(p) = \frac{N_{\alpha}(p) \Re}{\pi R_{D}^{2} \mu v} \frac{X(p)}{1 + \frac{X(p)}{X_{\alpha}}},
\end{equation}
where $\mu = 2 h n_{d} m_{p}$ is the surface density of the Galactic disc and $X_{\alpha} = m_{p}/\sigma_{sp,\alpha}$. Written in this form, the solution is very clear. At particle momenta for which the grammage is small compared with $X_{\alpha}$ the standard solution is recovered:
\begin{equation}
f_{0,\alpha}(p) = \frac{N_{\alpha}(p) \Re}{2 \pi R_{D}^{2} H} \frac{H^{2}}{D(p)},
\end{equation}
while in the range of momenta at which the solution is spallation-dominated, the observed spectrum reproduces the shape of the injection spectrum. 
In order to calculate the spectra of CRs at the Earth, an evaluation of the grammage is needed. In Ref. \cite{Ptuskin09}, the authors develop a leaky-box model in order to reproduce the GALPROP results and{, for a source spectrum $E^{-2}$ find a grammage in the form: 
\begin{equation}
X(R)=19\beta^{3}\left(\frac{R}{3GV}\right)^{-0.6}\rm{~g\,~cm^{-2}},
\label{eq:grammage_Pt}
\end{equation}
where $R$ is the rigidity of particles and $\beta=v/c\sim1$ for 1 TeV particles. They calculate also the grammage considering the reacceleration and find:
\begin{equation}
X(R)=7.2\left(\frac{R}{3GV}\right)^{-0.34}\rm{~g\,~cm^{-2}}.
\label{eq:grammage_Pt_reac}
\end{equation}
Given that the CR spectrum observed at Earth has an energy dependence $E^{- 2.65}$ as taken from TRACER and CREAM data \cite{ave08,yoon11}, in our calculations we used a slope of the diffusion coefficient $\delta=2.65-p_{inj}$, where for $p_{inj}$ we considered the values 2 and 2.31 as we will discuss in Sec~\ref{sec:results}. In the first case, the diffusion coefficient has a slightly stronger energy dependence than the above mentioned $E^{0.6}$; in the other case, instead, we use the exact form of Eq.~\ref{eq:grammage_Pt_reac}.}
For the spallation cross section we adopt the simple formulation of Ref. \cite{horandel07}:
\begin{equation}
\sigma_{sp}(E)[mb]= \alpha(E) A^{\beta(E)},
\label{sigmasp}
\end{equation}
where $A$ is the atomic mass number, and the two energy dependent coefficients are
$$\alpha(E)=5.44-7.93\,ln\left(\frac{E}{eV}\right)+0.61\,ln^{2}\left(\frac{E}{eV}\right)$$
$$\beta(E)=0.97-0.022\,ln\left(\frac{E}{eV}\right).$$

It is worth noticing that one of the values of $\delta$ we adopt here, and in fact the one that leads to most interesting results, is rather large, $\delta=0.65$ (corresponding to $p_{inj}=2$), and this is known to lead to exceedingly large anisotropy in the CR signal at Earth \cite{ptuskin06,blasiamato12b}.

\section{Results}
\label{sec:results}
Having shown that type II SNe are the most likely sources to accelerate protons up to the \textit{knee}, in this section we focus on this type of explosions and attempt at fitting the all particle spectrum. To this aim we vary the different parameters, including the ejecta density profile.

As discussed in \S~\ref{sec:NHR}, the maximum energy up to which CRs can be accelerated in a SN is regulated by the escape of particles from the shock region and in the very early stages of the SN evolution, in principle, this energy can exceed the \textit{knee} energy. However, the amount of mass processed at such times is relatively small and the net flux of particles is correspondingly small. The spectrum of accelerated particles that escape the shock region during the Sedov-Taylor phase and become CRs reflects the spectrum in the source if $f_s\propto E^{-(2+\beta)}$ with $\beta\geq 0$, and is $\propto E^{-2}$ if $\beta \le 0$. This spectral slope extends up to what we can name the effective maximum energy, $E_{M}$, reached at the end of the ejecta dominated phase of expansion. We find that, at energies larger than $E_{M}$, the spectrum is not falling exponentially, but rather as a power-law with a steeper index as compared to the lower energy trend. After escape from individual sources, CRs propagate diffusively through the Galaxy.  At the high energies we are interested in here, the only relevant process during propagation is spallation, while other processes, such as reacceleration or Galactic winds, induce negligible effects and are therefore ignored in the present calculation. It is also worth stressing that the particles that are actually accelerated in the early phases of the SN explosion (say, within the first 100 years) are the highest energy particles; consequently, no direct constraint on these accelerators can be inferred from $^{59}Ni$, which is often used to impose a lower limit on the time lag between the explosion and the acceleration phase. In any case, in this scenario, the material from which particles are accelerated to the highest energies is not polluted with SN ejecta, since it is basically the material expelled during the RSG phase of the pre-supernova star.

\begin{figure}[!h]
\includegraphics[scale=.8]{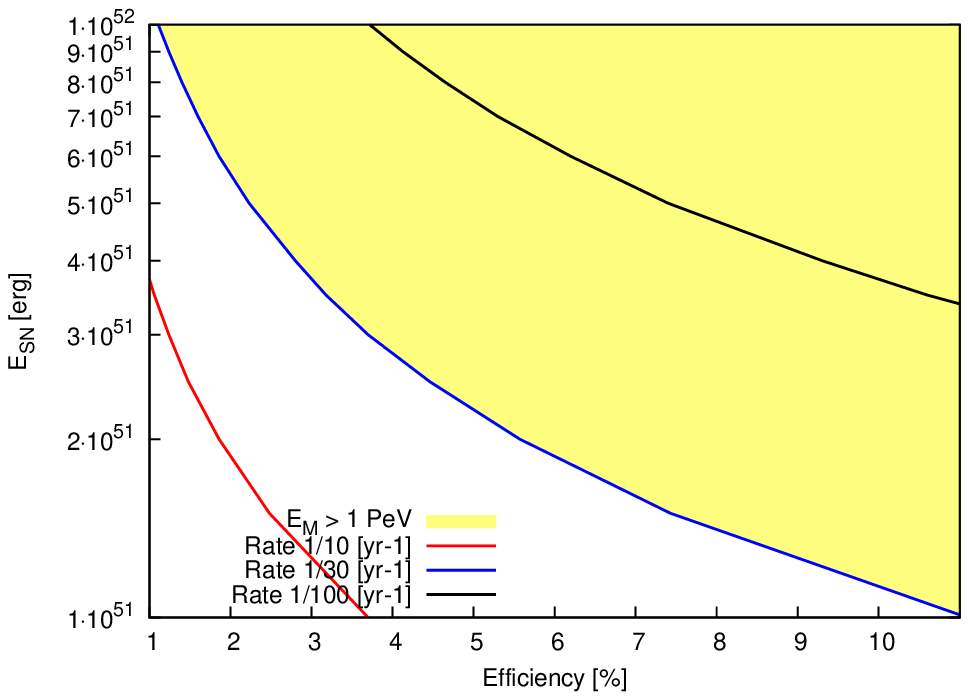}
\includegraphics[scale=.8]{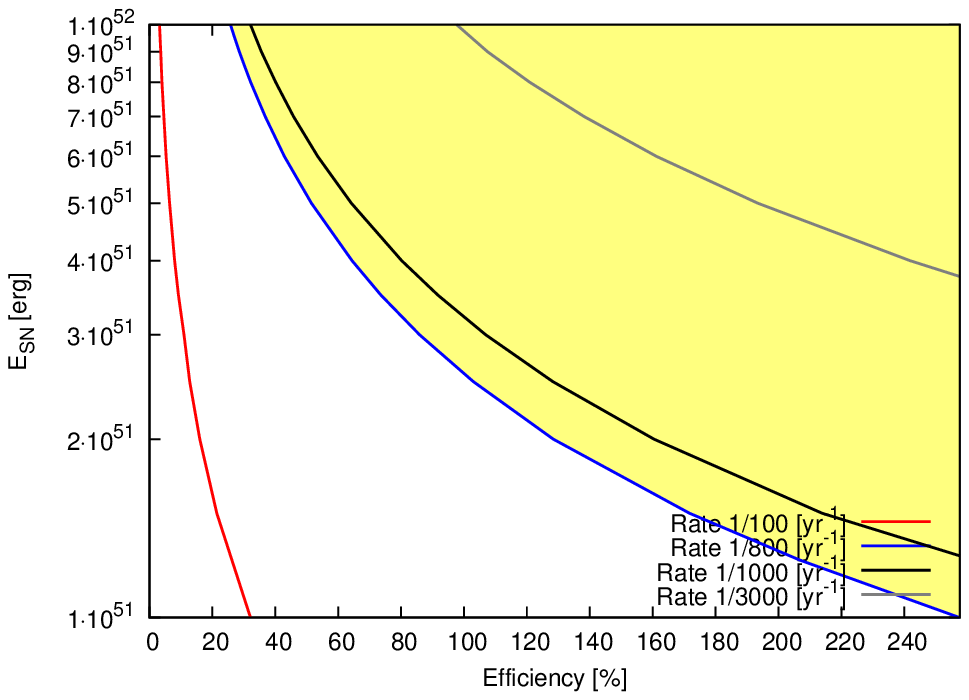}
\caption{\textbf{Left Panel}: injection index $p=-2$. \textbf{Right Panel}: injection index $p=-2.31$. The shaded area in the $\xi_{CR}-E_{SN}$ plane indicates the allowed range of parameters to reach $E_{M}=1$ PeV. The lines indicate the combination of values of the parameters for which the observed proton flux is also fitted. Each line refers to a given SN rate (as labeled).}
\label{fig:pars}
\end{figure}

As discussed in \S~\ref{sec:NHR}, the value of $E_{M}$ is determined by the combination of the CR acceleration efficiency, $\xi_{CR}$, and the energetics of the SN, $E_{SN}$. On the other hand, the flux of CRs at the Earth derives from a combination of $\xi_{CR}$, $E_{SN}$ and the rate of SNe, $\Re$. The shaded (yellow) region in Fig.~\ref{fig:pars} shows the region of the $\xi_{CR}-E_{SN}$ plane for which $E_{M}\geq 1$ PeV. The two panels refer to two different spectra at the source: $E^{-2}$ is on the left and $E^{-2.31}$ on the right. 

The need to fit the proton spectrum in the \textit{knee} region imposes an additional constraint on the combination $\xi_{CR}\,E_{SN}\,\Re$. In Fig.~\ref{fig:pars} the lines indicate such a constraint projected on the $\xi_{CR}-E_{SN}$ plane for SN rates as specified in each panel.

One can see that for the case of a "flat" source spectrum $E^{-(2+\beta)}$ with $\beta\le 0$, the standard SN energetics of $10^{51}$ erg is compatible with the need to have maximum energy in the PeV range with totally reasonable values of the CR acceleration efficiency ($\xi_{CR}\sim 10\%$) and SN rate ($\Re\sim 1/30\,\rm{yr^{-1}}$). If the rate is decreased to $1/100\,\rm{yr^{-1}}$, higher energetics and higher efficiencies are required to reproduce the CR flux at Earth and still accelerate particles up to 1 PeV, whereas for a SN rate $\Re\ge 1/30\,\rm{yr^{-1}}$, the \textit{knee} simply cannot be reached: this is because for large SN rates the allowed CR energetics is low in order to avoid overproducing the flux at Earth. The conclusion is even more severe for SNRs with a steeper internal spectrum: in this case, the sources that can reach the \textit{knee} and fit the spectrum are extremely rare, with a frequency less than $1/800\,\rm{yr^{-1}}$; in addition the source energetics is required to exceed a few $\times 10^{51}\,\rm{erg}$ and the acceleration efficiency several tens of percent.

\begin{figure}[!h]
\centering
\includegraphics[scale=1]{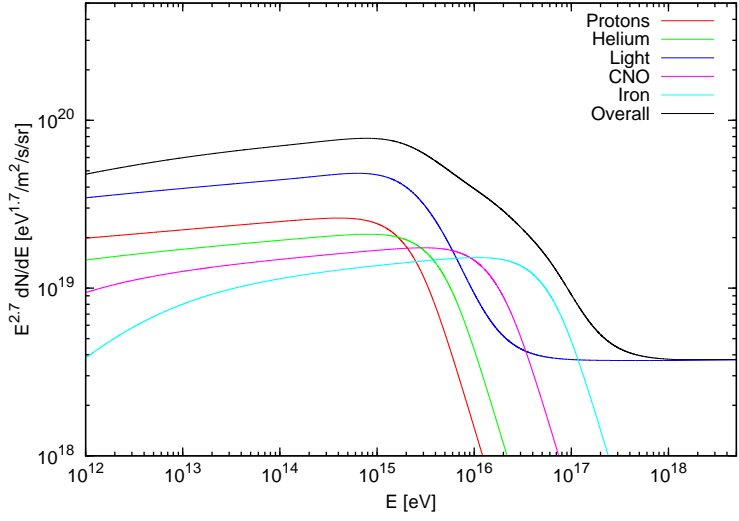} 
\caption{CR spectrum obtained with our model, fixing $k=9$, $E_{SN}=10^{51}$ erg and $\Re=1/30$ yr$^{-1}$. Protons(red), Helium (green), CNO (magenta), Iron (cyan), light component, p+He (blue) and the overall spectrum (black). At the highest energies we added by hand a constant extragalactic component. The proton \textit{knee} is at $E_{M}^{p}=1$ PeV, and the \textit{knee} of the other elements is at $E_{M}^{Z}=Z\times E_{M}^{p}$.}
\label{fig:k9}
\end{figure}

The conclusion is that the most common type II SNe, expanding in their red supergiant wind, with standard energetics and an efficiency of order 10\% are the most likely contributors of CRs up to the \textit{knee}.
In Fig.~\ref{fig:k9} we show the spectra of the different nuclei, as well as the all-particle spectrum, for a population of type II SNe with a density profile of the ejecta described by $k=9$, an energy release corresponding to $E_{SN}=10^{51}$ erg, exploding at a rate $\Re=1/30$ yr$^{-1}$. The spectra of nuclei are normalised to the data of CREAM for protons \cite{yoon11}, PAMELA for Helium \cite{adriani11} and TRACER for CNO and Iron \cite{ave08}. The diffusion coefficient is taken as described in Sec.~\ref{sec:observed}. Clearly, in a simple approach such as the one discussed here, it is not possible to account for subtleties such as the harder spectrum of helium compared to hydrogen. We simply normalize the abundances in such a way that the CR composition at Earth is reproduced. We consider this assumption as a weak point of all models trying to explain the origin of CRs: there is still no comprehensive theory of acceleration of nuclei in SN shocks. In fact, it is known, and can be easily understood, that there is a preferential injection of heavier nuclei at a collisionless shock \cite{Eichler}, but a quantitative theory of this phenomenon is not available yet. The situation is made even more complex by the fact that most refractory elements are embedded in dust grains, so that dust sputtering at shocks is bound to have en important role in determining chemical composition of CRs observed at the Earth. At present, the last and most comprehensive work on nuclei injection from dust sputtering is the one of Refs.~\cite{Meyer98,Drury98b}, but a physical understanding, from first principles, of injection of particles with different masses and charges, is still lacking, so that one is forced to parametrize the injection of nuclei: this is what is commonly done and what we have done in this work, fixing the abundances so as to fit observations.

Fig.~\ref{fig:k9} shows that for the values of the parameters that are typical of type II SNe the maximum energy of protons is in the PeV range, close to the \textit{knee} at $3\times10^{15}$ eV. The maximum energy of heavier nuclei is $Z$ times larger. The spectrum of protons (as well as of other nuclei) is not cut off sharply at the highest achievable energy, but rather shows a severe steepening to a different power law. The slope of the high energy part is determined by the profile of the ejecta. For $k=9$ the spectrum at the source is steepened from $E^{-2}$ to $E^{-4}$. The low energy part of the spectra of nuclei heavier than He shows a hardening that is due to spallation during propagation. The superposition of the spectra of all nuclei returns the all particle spectrum (black curve) that shows clear evidence for a \textit{knee}, at energy of $\sim 1$ PeV. The high energy hardening of the protons spectrum is due to the introduction of an {\it ad hoc} extragalactic proton component with spectrum $\sim E^{-2.7}$, as found by the KASCADE-Grande experiment.

\begin{figure}[!h]
\centering
\includegraphics[scale=1]{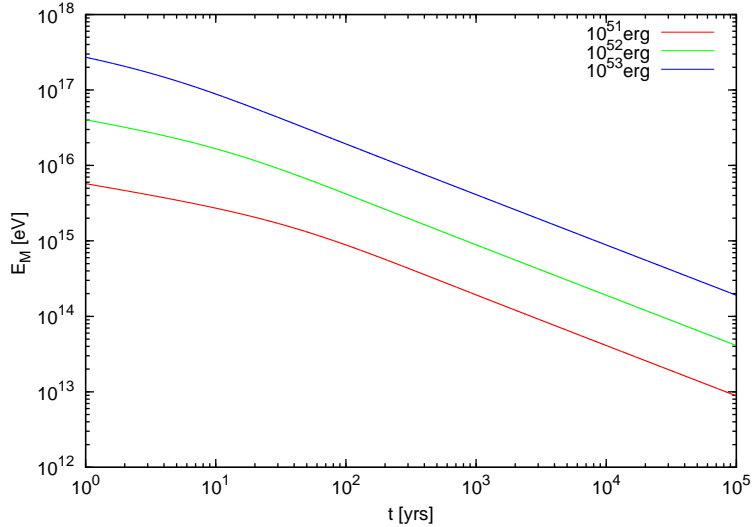}  
\caption{The maximum energy as a function of time after SN explosion with for different values of the total SN energy $E_{SN}$. The change of slope identifies the transition from ejecta dominated phase to Sedov-Taylor phase.}
\label{fig:times}
\end{figure}

In Fig.~\ref{fig:times} we plot the maximum energy of accelerated particles as a function of time for an acceleration efficiency of $10\%$ and different energetics of the parent SN, as labeled. One can see that more energetic SNe accelerate particles to higher energies and the onset of the Sedov phase in the wind occurs at earlier times. In any case, one should appreciate how in the framework of particle acceleration in SNe that explode in the wind of the pre-supernova star, the most important phase in the acceleration process occurs a few tens of years after the explosion, thereby implying a change of paradigm in which SNe, rather than SNRs, play a central role. 
The fact that in principle increasingly larger energies are reached at earlier times raises the issue of what is the minimum time that we need to take into account in the calculation of the spectra of escaping particles. One obvious limitation comes from the fact that, after escape, particles need to cross the wind thereby suffering different kinds of energy losses. For protons, inelastic $pp$ scattering occurs with a loss time scale $\tau_{pp}=1/n \sigma c$, and it is unimportant when
$$
\frac{4\pi r^{2} v_{w} m_{p}}{\dot M \sigma_{pp} c} \gg \frac{r}{c} \Rightarrow r\gg 10^{12} \rm cm 
$$
where reference values of the parameters have been adopted. For a shock velocity of $10^{4}$ km/s, this corresponds to times $t\gg 10^{3}$ s. 
Given the large photon background in the early phases of the SN explosion, it is worth considering also inverse Compton scattering losses of protons, with rate of energy change:
\begin{equation}
\frac{dE}{dt}=\frac{4}{3}\frac{\sigma_{t}}{\left(m_{p}/m_{e}\right)^{2}}c\left(\frac{E}{m_{p} c^{2}} \right)^2U_{ph}=4\times 10^{-17}t_{yr}^{-3} E_{GeV}^{2}\,\rm{GeV/s}
\end{equation}
where we approximate the photon energy density as $U_{ph}=E_{SN}/(\frac{4\pi}{3} R^{3})$. By requiring that $E/(dE/dt)\gg r/c$ one gets that $t\gg 6.3\times 10^{-6} E_{GeV}^{1/2}$ yrs. For protons with PeV energy this constrains times useful for escape to later than the first $\sim 2\times 10^{5}$ s. For more energetic SNe (namely larger values of the shock velocity) this bound becomes less constraining. 

Photodisintegration of nuclei in the dense photon background may also become important. However, this is a threshold process, which is turned on when $\epsilon \gamma_{A} > E_{bind}\sim 10$ MeV, $E_{bind}$ being the binding energy of a nucleon inside the nucleus. This leads to a lower bound on the nucleus Lorentz factor $\gamma_{A}>10^{7}$. Above this threshold, the time scale for the process to become irrelevant, assuming a typical photon energy in the SN of $\epsilon\sim 1$ eV, is
\begin{equation}
\frac{1}{n_{ph} \sigma_{ph} c} \gg \frac{r}{c} \to t\gg 80 ~\rm yrs.
\end{equation}
The bound becomes less severe for more energetic SNe, namely for larger shock velocities. We stress once more that this bound is of some concern only for nuclei with rigidity quite above the rigidity of the \textit{knee}. 

\subsection{Comparison with high energy data}
\label{sec:data}

Any understanding of the origin of CRs at the \textit{knee} is bound to have important implications for higher energy CRs, and more specifically for the transition from Galactic to extragalactic CRs. In this section we discuss this issue in some detail. The standard lore about the origin of the \textit{knee} in the all-particle spectrum is based on the premise that this feature coincides with a change of mass composition of CRs from light to heavy. In this perspective, the \textit{knee} is associated with the {\it end} of the spectrum of the light component of CRs, say H and He. The natural implication of this line of thought is that the sources of Galactic CRs must be able to accelerate particles to at least the energy of the \textit{knee} in the all-particle spectrum, say $\sim 3$ PeV. The important conclusion has stimulated endless discussion for the last thirty years on the sources and on the type of acceleration mechanism: in 1983, two seminal papers \cite{lagage83a,lagage83b} reached the conclusion that even in the presence of resonant growth of Alfv\'en waves the maximum energy reachable in SNRs could not exceed $\sim 10^{3}-10^{5}$ GeV, thereby falling short of the \textit{knee} by more than one order of magnitude. Measurement of the proton and He spectrum by KASCADE \cite{apel09} seemed to confirm that the spectrum of the light CR component is steepening in the PeV region. A note of caution should be added, in that the information about mass composition is accessible to KASCADE only through a complex Monte Carlo dependent procedure. For instance using Sybill and QGSJet to simulate the showers, would lead to different spectra inferred for H and He. 

More recently, the KASCADE-Grande experiment \cite{apel13} has measured the spectra and mass composition in the energy region $E\geq 10^{16.6}$ eV: the spectrum of protons was claimed to show an-ankle like feature at energies $E=10^{17.08 \pm 0.08}$ eV, where there is a hardening to a slope of about 2.7. This harder, high energy component was associated with the beginning of the extragalactic CR component. At the same time, the spectrum of Fe nuclei was observed to have a \textit{knee}-like feature at $E = 10^{16.92\pm 0.04}$ eV, which also happens to be at $\sim 26$ times the energy of the \textit{knee}. This feature was readily interpreted as the sign of a rigidity dependent \textit{knee} in the spectra of the different chemicals. Notice however that the high energy part of the spectra (above the \textit{knee}) does not appear to fall as an exponential, but rather as a steep power law. 

\begin{figure}[!ht]
\centering
\includegraphics[scale=0.7]{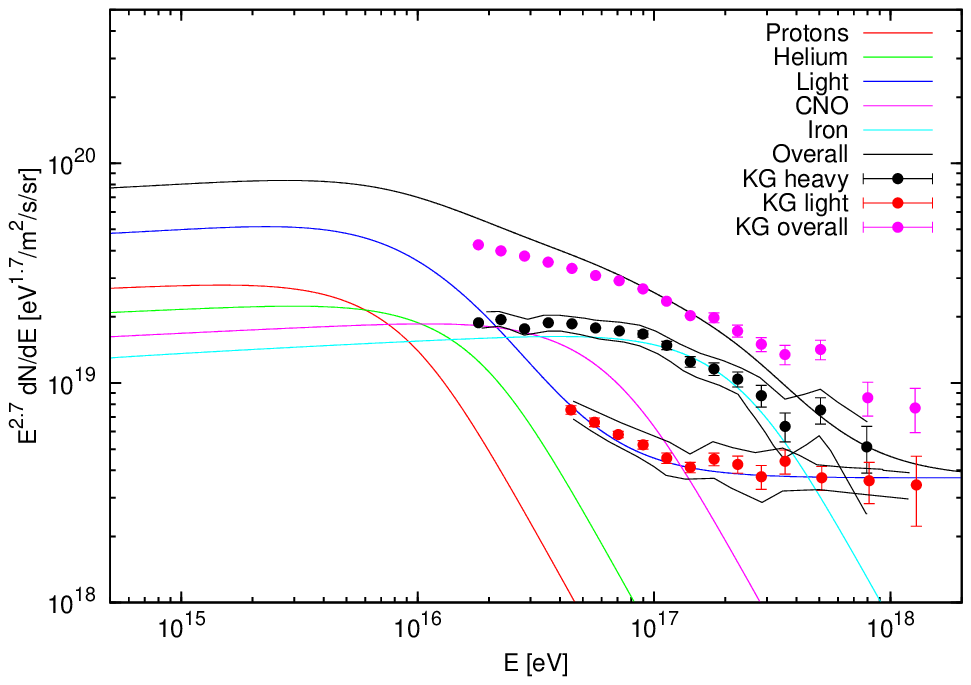}  
\includegraphics[scale=0.7]{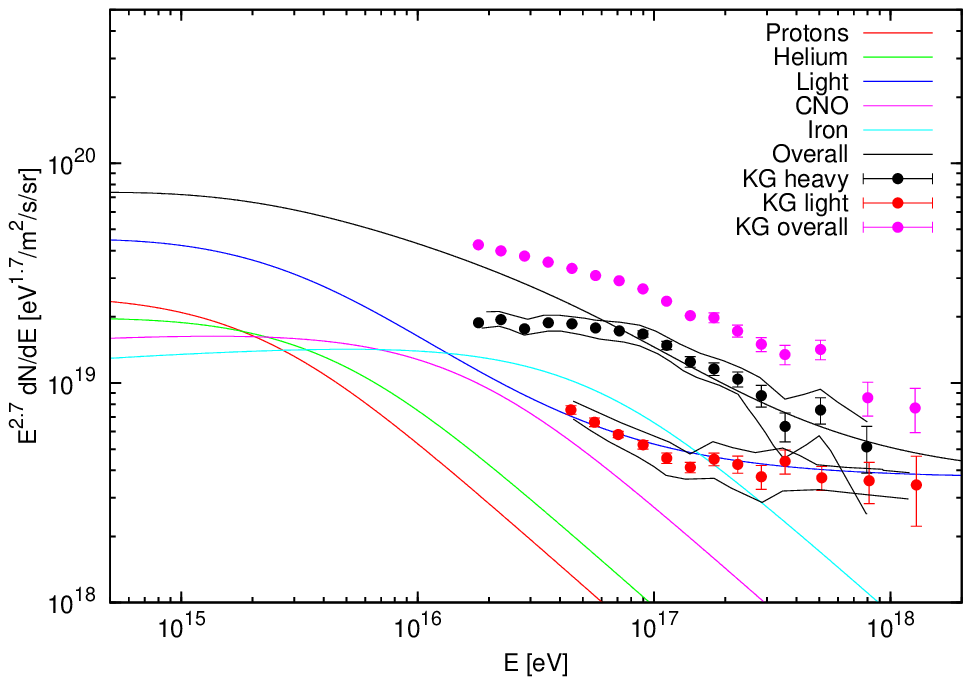}   
\caption{{\it Left)} our best fit model for KASCADE-Grande data \cite{apel13}, with $k=9$, $E_{SN}=2\times10^{51}$ erg, $\Re=1/110~yr^{-1}$, $E_{M}^{H}\cong 3.7$ PeV and $\xi_{CR}\cong20\%$. The highest energy data are fitted with an "ad hoc" extragalactic component. {\it Right)} Our best fit model for KASCADE-Grande data \cite{apel13}, with $k=7$, $E_{SN}=10^{51}$ erg, $\Re=1/25~yr^{-1}$, $E_{M}^{H}\cong 781$ TeV and $\xi_{CR}\cong12\%$.}
\label{fig:dataKG}
\end{figure}

Qualitatively, this behavior seems similar to the one discussed above and connected with the transition from ejecta dominated to Sedov expansion of a SN shell. 
In Fig. \ref{fig:dataKG} we show the spectra of different elements as calculated using the model discussed above, and compared with the KASCADE-Grande data on protons and Fe. An extragalactic proton component with power law shape $E^{-2.7}$ has been added to fit the highest energy data points of KASCADE-Grande. 
For the case $k=9$, most appropriate for type II SNe (left panel in Fig. \ref{fig:dataKG}), a maximum energy $E_{M}\cong 3.7$ PeV is required. Reaching this high energy requires a SN energy $E_{SN}=2\times10^{51}$erg, a factor of a few larger than the standard $E_{SN}=10^{51}$erg, an efficiency $\xi_{CR}=20\%$ and a rate of explosion of $\Re=1/110~\rm{yr^{-1}}$. The value of $E_{M}$ is constrained by the need to fit the ankle-like feature on the proton spectrum as measured by KASCADE-Grande. For $k=7$, less justified for this type of SNe, the high energy part of the injected spectra are somewhat harder and match the observations better. This is illustrated in the right panel of Fig.~\ref{fig:dataKG}, where $E_{M}\cong 781$ TeV and $\Re=1/25~\rm{yr^{-1}}$.

Recent measurements of the spectrum of the light nuclei and the all-particle spectrum carried out with the ARGO experiment have raised serious doubts about the very foundations of the idea of light nuclei being accelerated to $\sim PeV$ energies \cite{argo,disciascio14}. ARGO finds a \textit{knee} in the light component (H+He) at $\sim 650$ TeV, while the all-particle spectrum is shown to have a \textit{knee} at an energy compatible with all other measurements. This result appears to be compatible with some previous results obtained by experiments built at altitudes close to the shower maximum. The discrepancy between the ARGO and KASCADE results is an intrinsically observational issue and may be suggesting a rather poor knowledge of the development of showers and question our way to infer mass composition from shower-related observables. 

\begin{figure}[!ht]
\centering 
\includegraphics[scale=1]{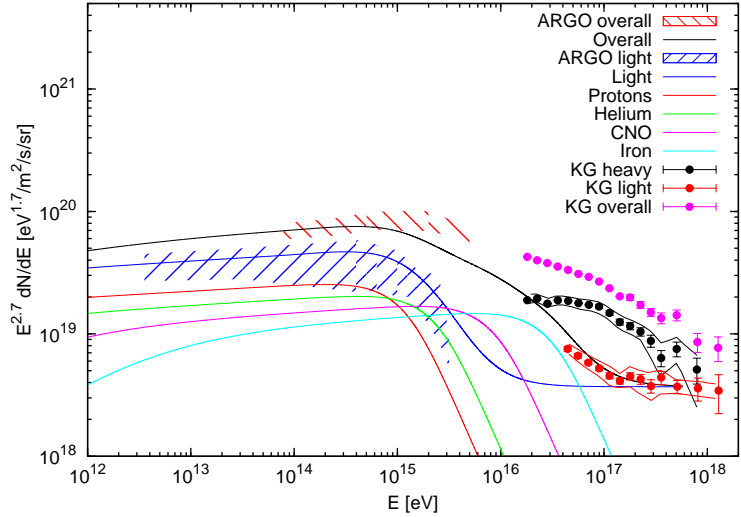} 
\caption{Best fit model to ARGO data \cite{argo,disciascio14}, with $k=9$, $E_{SN}=10^{51}$ erg, $\Re=1/15~yr^{-1}$, $E_{M}\cong 507$ TeV and $\xi_{CR}\cong5.2\%$.}
 \label{fig:dataARGO}
\end{figure}

Since it is hard to imagine a way to make ARGO and KASCADE data compatible with each other, we decided to take ARGO data at face value and tried to infer the consequences that would follow. This is illustrated in Fig.~\ref{fig:dataARGO}, where we plot the results of our calculations and compare them with the ARGO data, shown as a shadowed area. The relatively large area derives from the fact that a few different analysis techniques have been adopted so as to establish the independence on the Monte Carlo procedure adopted of the \textit{knee}-like feature in the light component. 

The best fit parameters we find in order to reproduce the ARGO data with our model are $E_{SN}=10^{51}$ erg, $\Re=1/15~yr^{-1}$, $E_{M}\cong 507$ TeV and $\xi_{CR}\cong5.2\%$. While it is clear that the spectrum of light nuclei (H+He) is easily accounted for, it is equally clear that the all-particle spectrum \textit{knee} cannot be reproduced within a simple approach such as the one discussed here. 

\section{Summary}
\label{sec:summary}
Most of the models for the origin of Galactic CRs still rely on SNRs as being the most likely sources. Yet, the issue of whether SNRs are able to accelerate CR protons up to the energy of the \textit{knee}, a few PeV, remains open. From the point of view of energetics, it seems that SNRs, in one flavour or another, are realistic candidates. From the point of view of the chemical composition, the anomalous ratio of $^{22}Ne/{^{20}}Ne$ observed in CRs has led to suggest that the bulk of Galactic CRs could be accelerated in OB associations \cite{higdon98,higdon03,binns08}, although this argument has been recently revised by \cite{prantzos12} who reached the conclusion that OB associations cannot contribute the majority of CRs.

In the present paper we focused on the issue of reaching the energy of the knee in particle acceleration during the expansion of a SN shock in the dense region occupied by the wind of a RSG, and stressing the physical aspects of the acceleration process, namely the nature of the instabilities that are required to bring the magnetic field strength and topology to the level that is suitable for particle acceleration. In particular, we investigated in detail the implications of the so called Non-Resonant Hybrid (NRH) instability described by \cite{bell04,schure13,bell13} by computing the maximum energy and the overall particle spectrum produced during the whole SN expansion, both in the ED and ST phases. The general idea is that at any given time, the particles that escape the system with the highest available energy $E$ produce the turbulence needed to scatter the next generation of particles with the same energy, thereby leading to acceleration to higher energies. We discussed the role of the NRH instability in both type Ia and type II SNe, in the context of three models of particle acceleration at the SN shock: 1) a flat spectrum $\propto E^{-2}$; 2) a hard spectrum $\propto E^{-(2+\beta)}$ with $\beta<0$, and 3) a steep spectrum $\propto E^{-(2+\beta)}$ with $\beta>0$. The maximum energy and the spectrum of escaping particles was calculated in each case. 

The effective maximum energy, which is reached at the beginning of the ST phase in all cases, is found to be at most of order a few hundred TeV for type Ia SNe exploding in the ISM (with the typical values of the relevant parameters). For type II SNe, the situation is quite more complex because these explosions often occur in the wind environment created by the pre-supernova red-giant star. Particle acceleration during the expansion of the SN shock in the dense slow wind of the RSG progenitor is found particularly promising from the point of view of exciting the NRH instability. In this case, the instantaneous maximum energy is shown to always decrease as a function of time after the explosion. However, the ST phase typically starts a few tens of years after the SN event and most of the mass is processed by the end of the ejecta dominated phase. These facts have important consequences. The early beginning of the ST phase allows for a maximum energy which is still high; the effective maximum energy does not coincide with a cut-off, but rather with a steepening of the spectrum, since higher energy particles are indeed produced during the ED phase. Our calculations show that a type II SN with standard energetics ($E_{SN}=10^{51}$ erg, $\xi_{CR}=10\%$) can accelerate CRs up to \textit{knee} energies, $E_{M}\sim1$ PeV. 

It is interesting to appreciate how particle acceleration at PeV energies typically occurs $\sim 10-30$ years after the SN explosion, which represents a change of paradigm with respect to the standard SNR paradigm where the highest energies are reached  several hundred years after explosion. In this scenario, the probability of catching a PeVatron in action in our own Galaxy (for instance through its gamma ray emission) is very low.

For given parameters of the SN explosion, the maximum energy reached at the beginning of the ST phase is a function of the CR acceleration efficiency while the spectrum observed at the Earth also depends on the rate of SN explosions. We calculated the required efficiency and rates necessary to reach the \textit{knee} and fit the overall CR flux: for steep spectra at the shock, the escape current is lower and this forces one to have larger SN energetics and larger CR acceleration efficiencies, which is counterintuitive for a steep spectrum of accelerated particles. For flat $E^{-2}$ spectra at the shock, the spectrum of escaping particles is also flat and PeV energies can be reached for ordinary parameters of the SN explosion. Reaching PeV energies is even easier for hard spectra at the shock, however in both the flat and the hard case, one has to face the well known severe problem with CR anisotropy \cite{ptuskin06,blasiamato12b,ptuskin12}. 

The comparison of the expected spectra of light nuclei with the recent data of KASCADE-Grande and ARGO appears very problematic: while individually both sets of data can be fitted with reasonable values of the SN parameters and CR acceleration efficiencies, there is no model that can fit both at the same time. KASCADE-Grande data require a SN energy $E_{SN}=2\times10^{51}$erg, an efficiency $\xi_{CR}=20\%$ and a rate of explosion of $\Re=1/110~\rm{yr^{-1}}$ that lead to a maximum energy $E_{M}\cong 3.7$ PeV, whereas we can fit ARGO data with $E_{SN}=10^{51}$ erg, $\Re=1/15~yr^{-1}$ and $\xi_{CR}\cong5.2\%$ that lead to $E_{M}\cong 507$ TeV. The disagreement between these two experiments is a purely experimental problem, that requires a serious and careful assessment of the systematic uncertainties involved in the adopted experimental techniques.

\section*{Acknowledgments}
We thank I. De Mitri for several discussions on ARGO data and M. Bertaina for discussion of the KASCADE-Grande data. This work was partially funded through grant PRIN-INAF 2013.

\end{document}